# Constrained crystals deep convolutional generative adversarial network for the inverse design of crystal structures

Teng Long*, Nuno M. Fortunato, Ingo Opahle, Yixuan Zhang, Ilias Samathrakis, Chen Shen,

Oliver Gutfleisch, Hongbin Zhang*

*Institute of Materials Science, Technische Universität Darmstadt, Darmstadt 64287, Germany*

**Correspondence to: tenglong@sdu.edu.cn / hongbin.zhang@tu-darmstadt.de*

**Autonomous materials discovery with desired properties is one of the ultimate goals for materials science, and the current studies have been focusing mostly on high-throughput screening based on density functional theory calculations and forward modelling of physical properties using machine learning. Applying the deep learning techniques, we have developed a generative model which can predict distinct stable crystal structures by optimizing the formation energy in the latent space. It is demonstrated that the optimization of physical properties can be integrated into the generative model as on-top screening or backwards propagator, both with their own advantages. Applying the generative models on the binary Bi-Se system reveals that distinct crystal structures can be obtained covering the whole composition range, and the phases on the convex hull can be reproduced after the generated structures are fully relaxed to the equilibrium. The method can be extended to multicomponent systems for multi-objective optimization, which paves the way to achieve the inverse design of materials with optimal properties.**

## Introduction

The discovery and exploitation of materials has enormous benefits for the welfare of society and technological revolutions [1], which motivates the launch of Materials Genome Initiative in 2011 [2,3]. Till now, high-throughput (HTP) workflows based on density functional theory (DFT) enable massive calculations on existing and hypothetical compounds, accelerating materials discovery dramatically [4]. For instance, crystal structure predictions can be performed based on brute force substitution of the known prototypes or the evolutionary algorithms as implemented in CALYPSO [5] and USPEX [6]. Nevertheless, the soaring computation cost prevents exhaustive screening over an immense phase space, limiting the application of such methods and calling for more efficient solutions. Global optimization methods added a target function to guide the searching and therefore reduce the computational cost [7], *e.g.*, crystal structure prediction with Bayesian optimization can efficiently select the most stable structure from a large number of candidate structures with a lower number of searching trials, and has been successfully applied to predict structures of NaCl and $Y_2Co_{17}$ [8]. The random search method has also demonstrated its potential to find the global minimum with highly parallel implementation, as manifested by the applications on elemental boron, silicon cluster and lithium hydrides [9,10]. Last but not least, the evolutionary algorithm is promising in predicting thermodynamically stable crystal structures, such as demonstrated for tungsten borides, carbon, and $TiO_2$ [11,12].

The emergent machine learning techniques have become the fourth paradigm for materials science [13], as exemplified by the rapid developments and applications in the last decade [14]. Particularly, the generative machine learning methods have been utilized to discover distinct materials, *e.g.,* applying neural networks to assist molecule design [15,16]. Two particularly powerful methods therein are the variational autoencoder (VAE) [17] and generative adversarial network (GAN) [18], resulting in significant progresses on molecule design [19]. Combined with the successful forward modeling of physical properties [20], generative machine learning on the existed structures enables inverse design, *i.e.*, prediction of distinct structures with desired properties [1,21]. Unfortunately, unlike molecules which can be represented using simplified molecular input line entry specification (SMILES) [22], inverse design of three-dimensional (3D) crystal structures has been rare due to the challenges in obtaining a continuous representation in the latent space [20].

Recently, Nouira *et al.* developed a CrystalGAN model to design ternary A-B-H phases starting from binary A-H and B-H structures using a vector-based representation [23], and successfully applied it on the Ni-Pd-H system. Another promising representation is the two-dimensional (2D) crystal graphs constructed based on the voxel and autoencoder methods [24–26]. For instance, Noh *et al.* developed the iMatGen model to represent the image-based crystals continuously, and successfully generated distinct $V_xO_y$ using VAE [27]. The ZeoGAN scheme developed by B. Kim *et al.* is based on the same representation applied on zeolites, where Wasserstein GAN (WGAN) is used to generate porous materials [28]. S. Kim *et al.* also developed the Crystal-WGAN to design Mn-Mg-O system using such a continuous representation [29]. These models exhibit excellent abilities in generating distinct structures, but not all predicted phases host the desired functionality. For example, only ZeoGAN integrates the optimization of heat of adsorption, whereas the other schemes exert mostly constraints to filter the generated structures. It is noted that by nature, inverse design entails the optimization of properties in the latent space, and thus internalizes the desired property as a joint objective of the generator [30].

In this work, we developed a GAN-based inverse design framework for crystal structure prediction with target properties and applied it on the binary Bi-Se system. It is firstly demonstrated that our deep convolutional generative adversarial network (DCGAN) can be applied to generate distinct crystal structures [31]. Taking formation energy as the target property, its optimization is integrated into the DCGAN model in two schemes: DCGAN + constraint to select structures following the conventional screening method, and constrained crystals deep convolutional generative adversarial network (CCDCGAN) with an extra feedback loop for automatic optimization. The performance of the DCGAN + constraint and CCDCGAN models are comparatively evaluated, and it is demonstrated CCDCGAN is more efficient in generating stable structures, while the DCGAN + constraint model has a higher generation rate of meta-stable structures. Both schemes can be generalized to perform multi-objective inverse design for multicomponent systems [32], hence our work paves the way to achieve autonomous inverse design of distinct crystal phases with desired properties.

## Results

### Data for generative adversarial network

Typically, DCGAN requires $10^3$ known crystal structures as positive examples during training. However, there are mostly tens of known crystalline phases for a given binary system, *e.g.*, Materials Projects (MP) [33] provides only 17 already known $Bi_xSe_y$ materials, which are not sufficient for deep learning. In order to get enough training data, DFT calculations are performed on the prototype binary structures used in Ref. 27, where are selected based on two criteria: (1) there are no more than 20 atoms in the unit cell and (2) the maximum lattice constant is smaller than 10 Angstrom. Starting from 10981 prototype structures, we managed to converge 9810 cases after substituting Bi and Se atoms [34], which are taken as training database. Hereafter, such a database is referred as the Bi-Se database. Out of the 15 experimentally achievable phases [35], 8 structures (*i.e.*, $BiSe_3$, $Bi_2Se_3$, $Bi_4Se_5$, $Bi_3Se_2$, $Bi_5Se_3$, $Bi_2Se$, $Bi_7Se_3$, and $Bi_3Se$) are included in the database, whereas the other 7 compounds (*i.e.*, $BiSe_2$, $Bi_3Se_4$, $Bi_8Se_9$, BiSe, $Bi_8Se_7$, $Bi_6Se_5$, and $Bi_4Se_3$) are excluded by the selection criteria, as demonstrated in Supplementary Figure 1. Furthermore, out of the Bi-Se database, 155 (1.56%) structures of the database are stable (*i.e.,* formation energy lower than 0 eV/atom and distance to the convex hull smaller than 100 meV/atom) and 707 (7.2%) cases are meta-stable (*i.e.,* with negative formation energy and distance to the convex hull smaller than 150 meV/atom).

### Continuous representation for crystal structures

Although the crystallographic and chemical information are stored in the crystallographic information file (CIF), the discontinuous and heterogeneous formats are not a suitable choice of representation in a generative model, thus a continuous and homogeneous representation including both the chemical and structural information is required. Following Ref. 27, the lattice constants and atomic positions are translated into the voxel space, followed by encoding into a 2D crystal graph through autoencoder, as demonstrated in Fig. 1(a). In this way, a continuous latent space is constructed. The whole process is reversible, *i.e.,* a random 2D crystal graph can be reconstructed into a crystal structure in real space, which is essential for a generative model. When applied to the Bi-Se database, 9420 of 9810 (*i.e.*, 96%) crystal structures are successfully reconstructed. Such a high ratio is not caused by overfitting, illustrated by the learning curve of both autoencoders as shown in Fig. 1 (b) and (c). The differences of training set loss and test set loss are both negligible, suggesting that the 2D crystal graphs in the latent space contain adequate information to reconstruct crystal structures. This is confirmed by the diversity of twelve typical 2D crystal graphs in the latent space (cf. Fig. 1(d) for four cases). Please refer to Methods as well as Supplementary Information for more details.

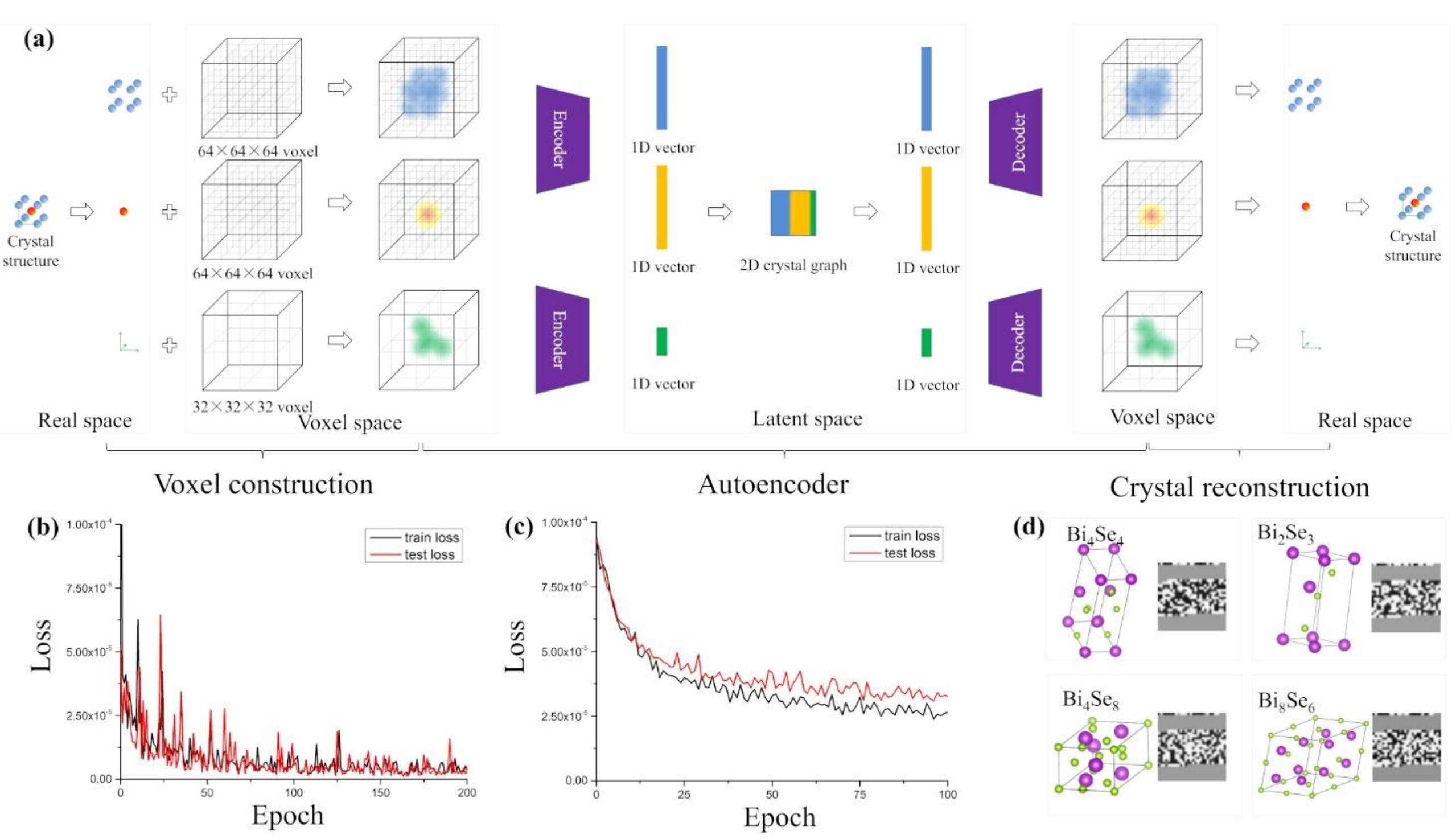

Fig. 1. Construction of 2D crystal graphs. (a) Schematic diagram of generating 2D crystal graphs; (b) Learning curve of lattice autoencoder, where the red line represents the test loss and the black line represents the training loss; (c) Learning curve of sites autoencoder, where the red line represents the test loss and the black line represents the training loss; (d) Four typical crystal structures ($Bi_4Se_4$, $Bi_2Se_3$, $Bi_4Se_8$, $Bi_8Se_6$) and their corresponding 2D crystal graphs.

## Construction and prediction of DCGAN

Turning now to the realization of generative models, VAE [17] and GAN [18] are two most popular algorithms. VAE is a mutation of autoencoder discussed above, which assumes a specific (such as Gaussian) distribution of data (in our case 2D crystal graphs) in the latent space. To be generative, such a distribution function should be defined properly, *i.e.,* to be consistent with the distribution of 2D crystal graphs of the known crystal structures in the latent space. In addition, the distribution function has to be specified prior to the training process and its form determines the performance of the generative model, which demands domain expertise in statistics and profound understanding of the input data. For instance, a recent work by Ren *et al.* used sigmoid function in VAE to design inorganic crystals and find structures that do not exist in training set [36]. Court *et al.* implemented VAE to generate crystal structures and successfully applied it to a broad class of materials including binary and ternary alloys, perovskite oxides, and Heusler compounds [37]. In contrast, GAN trains two mutual-competitive neural networks (*i.e.*, generator and discriminator) to generate data statistically the same as the training data without assuming the distribution function [18]. That is, the discriminator tries to distinguish the generated data from the training data, while the generator attempts to fool the discriminator by generating data similar to the training data. Mathematically the objective of GAN is defined by the following equation:

$$\max_{D}\left(\min_{G}\left(\frac{1}{2}\cdot E_{\mathbf{x}\sim p_x}\left[1-D(\mathbf{x})\right]+\frac{1}{2}\cdot E_{\mathbf{x}\sim p_g}\left[D(\mathbf{x})\right]\right)\right) \quad (1)$$

where $D$ is the discriminator, $G$ is the generator, $E$ means expectation value, $\mathbf{x}$ is the original data, $D(\mathbf{x})$ is the output of discriminator with $\mathbf{x}$ as input, $p_x$ is the possibility density function of the original data, while $p_g$ is the possibility density function of the generated data. Through the competition between the generator and discriminator, the distribution of generated data becomes hardly distinguishable with the distribution of training data. Compared with VAE, GAN does not require specifying a distribution function in the latent space ($p_g$), which makes the generation process more robust, and thus is adopted in this work [19].

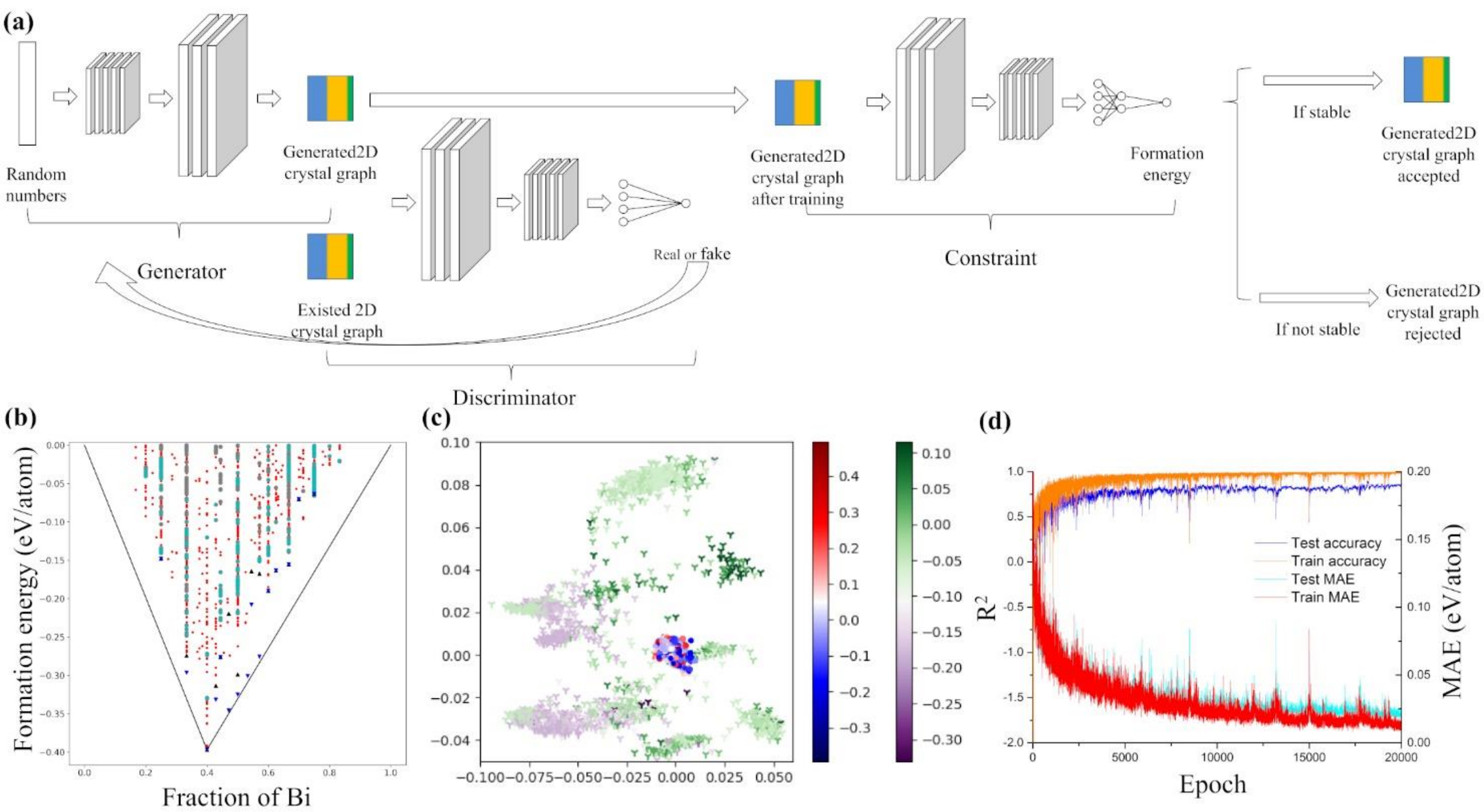

Fig. 2. Illustration on the DCGAN + constraint model. (a) Schematic diagram of the DCGAN + constraint model; (b) Scatter plot of formation energy vs. composition for structures generated by DCGAN and DCGAN+constraint models, where the black line represents the convex hull, red points denote the crystal structures considered in our machine learning database, gray circles indicate the generated structures, cyan stars denote the DCGAN+constraint structures, blue triangles indicate the experimentally achievable phases (cf. Supplementary Table 7), black triangles mark the crystal structures of the available Bi-Se phases in the Bi-Se database.; (c) Distribution of the crystal structures in the latent space, the circles denote the structures in the Bi-Se database, while the stars denote the generated structures. The color maps denote the corresponding formation energy; (d) Learning curves of the training process of the DCGAN model, red and orange represent the MAE and accuracy of training set, while blue and cyan represent the accuracy and MAE of the test set.

Such an implementation of DCGAN can generate crystal structures with a high success rate (defined as the ratio of the number of generated crystals over the number of generated 2D crystal graphs), *e.g.*, 2832 crystal structures are transformed from 13000 generated 2D crystal graphs. The generated structures cover a large composition range as shown in Fig. 2(b), where the red points denote the original data in the Bi-Se database and the gray circles mark the generated structures by DCGAN. One interesting observation is that the generated structures are mostly with negative formation energies, on average lower than that of the original database (Supplementary Figure 2(a)). For instance, 2548 structures out of 2832 have negative formation energies after follow-up DFT relaxations, which consists of 89.9% of the generated structures, whereas only 46.8% (4588 out of 9810) of the Bi-Se database have negative formation energies. Out of 2832 generated structures, it is found that 1233 of them are meta-stable, and 58 are stable.

Detailed analysis reveals that 476 out of the 2832 structures are considered to be *distinct* structures, *i.e.*, different from those in the Bi-Se database and different from each other, leading to 73 (15) meta-stable (stable) structures respectively. For instance, $Bi_2Se_4$ and $BiSe_3$ shown in Supplementary Figure 2(b) and (c) are the automatic generated structures, whose formation energies (distances to the convex hull) are -0.192 (0.136) eV/atom and -0.135 (0.111) eV/atom, respectively. Furthermore, DCGAN explores a much larger phase space when generating structures. As illustrated in Fig. 2(C), the structures in the Bi-Se database are concentrated in a small region in the latent space, whereas much larger phase space is covered by DCGAN. Thus, DCGAN can generate distinct crystal structures beyond the phase space of the known crystal structures.

**Training of constraint model**

As mentioned above, the objective of inverse design is to design compounds with desired properties, including thermodynamical, mechanical, and functional properties. In this work, we take the formation energy (*i.e.*, thermodynamic stability) as the target property. In order to be able to optimize the formation energy in the latent space, another convolutional neural network (CNN) model is trained taking 2D crystal graphs in the latent space as descriptors and formation energies as the output physical property. The training is carried out by a 90%/10% training/test-set ratio over the Bi-Se database, with the resulting $R^2$ score and mean absolute error (MAE) being about 85% and 0.019 eV/atom (Supplementary Figure 3(a)), respectively. Such a MAE is even smaller than that (0.021 eV/atom) obtained using the state-of-the-art CGCNN model [38]. That is, the 2D crystal graphs in the latent space can be considered as effective descriptors to perform forward prediction of physical properties. We note that overfitting is a marginal issue for such a high accuracy, as seen in the learning curve of the training process in Fig. 2(d). Details on the CNN model are described in Methods and Supplementary Information.

**Construction and prediction of DCGAN + constraint**

A straightforward way to implement the optimization of physical properties, *e.g.*, formation energy in this work, is to apply an add-on constraint on the generated structures using DCGAN, as sketched in Fig. 2(a). The advantage doing so is that there is no need to train another model, which saves training time. Additionally, all the existing machine learning models can be transplanted no matter whether such forward predictions can take 2D crystal graphs in the latent space or vector based chemical and structural information as descriptors. This allows the optimization of a wide spectrum of physical properties [20]. Nevertheless, this method is essentially a selection of the structures generated by DCGAN, thus it cannot search automatically for a specific region in the latent space to reach the local optimal values. More details are described in Supplementary Information.

The DCGAN+constraint model demonstrates its advantage in generating metastable structures. Applying the constraint on the formation energy (with a tolerance of 0.3 eV/atom) on 2832 generated structures by DCGAN, 2148 crystal structures are selected and optimized by further DFT calculations. Such a screening procedure does not affect the diversity of the composition as shown in Supplementary Figure 2(d). The application of such a constraint has evident effect in the latent space, as demonstrated in Supplementary Movie 1 where the constraint screens the unwanted points out leading to a shrunk region in the latent space. With such a constraint applied, the ratio of structures with negative formation energy is higher, increasing from 89.9% for DCGAN to 94.8% (2037/2148). The ratio of meta-stable structures also becomes higher compared to the DCGAN case, *i.e.*, 1145 meta-stable structures are generated from 2037 structures, reaching the ratio of 56.2% which is better than 43.5% in the DCGAN model. However, the number of generated stable structures reduces to 36 due to the selection process.

An interesting observation is that applying the constraint jeopardizes the generation of distinct structures. After performing DFT calculations, only 247 distinct structures remain, which are significantly reduced compared with the previous 476 cases obtained by DCGAN. The reason is two-fold. On the one hand, the generated crystal structures by DCGAN are not guaranteed to be at the mechanical and dynamical equilibrium, *i.e.,* the lattice constants and atomic positions will change during DFT relaxation. On the other hand, the CNN model applied as constraint is not good enough in predicting formation energy, though the MAE for the formation energy is only about 0.019 eV/atom. Thus, 629 candidates have been screened out by considering a tolerance of 0.3 eV/atom. After comparing with the known crystal structures in the Bi-Se database, 67 distinct meta-stable structures and 12 distinct stable structures are obtained, *e.g.*, two of them $Bi_3Se$ and $Bi_2Se_2$ are demonstrated in Supplementary Figure 2(e) and (f), with the formation energies (distances to the convex hull) being -0.065 (0.099) eV/atom and -0.202 (0.126) eV/atom, respectively. Overall, generated structures via DCGAN + constraint have a higher ratio of meta-stable structures.

**Construction and prediction of CCDCGAN**

The constraint can also be integrated into DCGAN as a back propagator, as illustrated as CCDCGAN in Fig. 3(a), to realize automated optimization in the latent space so that inverse design can be realized. In this way, the constraint gives rise to an additional optimization objective of the generator, which can be mathematically described as:

$$\min_{G}\left(\frac{1}{2}\cdot E_{\mathbf{z}\sim p_x}\left[D(\mathbf{z})\right]+\omega\cdot E_f\left(\mathbf{z}\right)\right),\mathbf{z}\sim p_g, \tag{2}$$

where $E_f$ is the formation energy predicted by the constraint model, **z** is the generated 2D crystal graph, and $\omega$ is defined as the weight of formation energy loss. Note that such an additional optimization objective cannot outweigh the primary objective, leading to lower weight for the formation energy loss (0.1 in this work) than the discriminator loss. Unlike the DCGAN + constraint model, CCDCGAN can accomplish automated searching for the local minima in the latent space and thus improve the efficiency of discovering distinct stable structures. It is noted that CCDCGAN requires a training from scratch, which takes 18 hours on a Quadro P2000 GPU machine, which is 2 hours longer than the time needed for training the DCGAN + constraint model. Please refer to Methods and Supplementary Information for more details.

In comparison with the other two models (*i.e.*, DCGAN and DCGAN+constraint models), CCDCGAN has a higher success rate of generating crystal structures, *i.e.*, 3743 crystal structures are transformed successfully from 13000 generated 2D crystal graphs. Correspondingly, the number of structures which are with negative formation energy and meta-stable is 3556 and 1910, respectively. Furthermore, CCDCGAN has higher generation efficiency of stable structure (307 out of 3743) when compared with the other two models, which suggests that the back propagation indeed causes optimization in the latent space. This is also proved by the average formation energy of generated structures, where the average formation energy (-0.11 eV/atom) for the CCDCGAN generated structures is lower than that from the DCGAN (-0.08 eV/atom) and DCGAN + constraint (-0.09 eV/atom).

The most intriguing result is that the number of distinct structures is bigger than that from DCGAN or DCGAN + constraint, where 511 distinct structures have been identified after performing DFT relaxations on 3743 structures. Correspondingly, the number of distinct meta-stable and stable structures is 187 and 42, respectively. Such numbers are similar to those obtained using iMatGen and Crystal-WGAN [27,29], indicating comparable performance of our DCGAN-based models with the existing ones. Three particularly promising structures with formation energy lower than their counterparts in Bi-Se database are $Bi_2Se_4$, $Bi_4Se_2$, and $Bi_6Se_2$, whose formation energies (distance to the convex hull) are -0.287 (0.041) eV/atom, -0.139 (0.069) eV/atom, and -0.063 (0.101) eV/atom, respectively. In this regard, CCDCGAN performs better than both the DCGAN + constraint and DCGAN models in generating distinct structures.

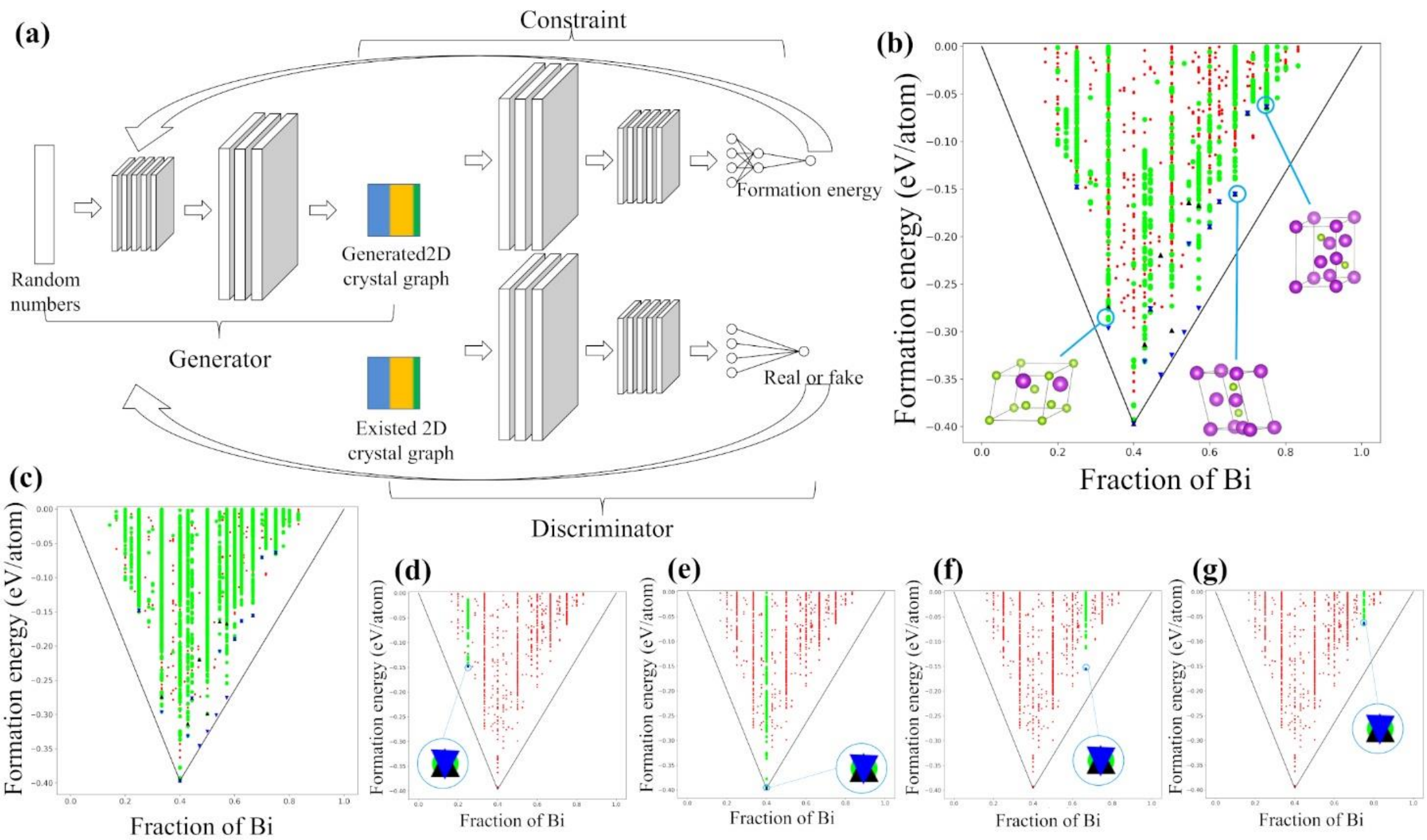

Fig. 3. Illustration on the CCDCGAN model. (a) Schematic diagram of the CCDCGAN model; (b) Scatter plot of formation energy vs. composition of CCDCGAN generated structures, where the black line represents the convex hull, red points denote the crystal structures considered in our machine learning database, green circles represent the generated structures, blue triangles indicate the experimentally achievable phases, black triangles are the crystal structures of the available Bi-Se phases in the Bi-Se database; and crystal structure of generated $Bi_2Se_4$, $Bi_4Se_2$, and $Bi_6Se_2$. (c) Scatter plot of formation energy vs. composition of 100,000 structures generated by CCDCGAN; (d) The $BiSe_3$ phases regenerated by CCDCGAN which are trained with the $BiSe_3$ composition removed; (e) The $Bi_2Se_3$ phases regenerated by CCDCGAN which are trained with the $Bi_2Se_3$ composition removed; (f) The $Bi_2Se$ phases regenerated by CCDCGAN which are trained with the $Bi_2Se$ composition removed; (g) The $Bi_3Se$ phases regenerated by CCDCGAN which are trained with the $Bi_3Se$ composition removed.

To explore the full competence of the CCDCGAN model, we generated 100,000 crystal structures and performed

DFT calculation on the generated structures. The CCDCGAN model is able to regenerate most (11 out of 15) experimentally achievable phases. As seen from Fig. 3(c) and Supplementary Table 7, all the 8 experimentally achievable phases in the Bi-Se database can be regenerated using our CCDCGAN model, *i.e.*, exactly the same crystal structures as the experimentally achievable phases of $BiSe_3$, $Bi_2Se_3$, $Bi_4Se_5$, $Bi_3Se_2$, $Bi_5Se_3$, $Bi_2Se$, $Bi_7Se_3$, and $Bi_3Se$ can be obtained. Furthermore, our CCDCGAN model can also generate three out of seven non-included experimentally achievable phases, namely, $BiSe_2$, $Bi_3Se_4$, and $Bi_6Se_5$, with exactly the same crystal structures as reported. Unfortunately, the other four phases, namely, $Bi_8Se_9$, BiSe, $Bi_8Se_7$, and $Bi_4Se_3$, out of seven non-included phases cannot be reproduced, where the generated structures of BiSe and $Bi_4Se_3$ have comparable formation energy as the experimentally achievable structures, while the formation energies of generated $Bi_8Se_9$ and $Bi_8Se_7$ structures are higher than those reported phases. It is suspected that this can be attributed to the constraints (*i.e.*, unit cell dimension and number of atoms) we applied to select the training database, which will be extended in the future.

To further test the predictive capability of the CCDCGAN model, we deliberately removed the crystal structures of 4 specific compositions (*i.e.*, $BiSe_3$, $Bi_2Se_3$, $Bi_2Se$, and $Bi_3Se$) from the training database, and observed that the experimental crystal structures of such four phases can still be successfully regenerated, as shown in Fig. 3 (d) to (g). This proves that the CCDCGAN model can generate crystal structures for unknown compositions, and thus it could accelerate the discovery of distinct crystal phases.

## Discussion

We have developed an inverse design framework CCDCGAN consisting of generator, discriminator and constraint and successfully applied it to design unreported crystal structures with low formation energy for the binary Bi-Se system. It is demonstrated that 2D crystal graphs can be used to construct a latent space with continuous representation of the known crystal structures, which serve as effective descriptors for modeling the physical properties (*e.g.*, formation energy) and can be decoded into real space crystal structures enabling the generation of distinct crystal structures. Such an inverse design model can be easily generalized to the multicomponent cases, as demonstrated by Ref. 29. Importantly, we elucidate that the optimization of physical properties (*e.g.*, formation energy) can be integrated into the generative deep learning model as explicit constraints or back propagators. This allows further development of a multi-objective inverse design framework to optimize other physical properties by modifying the objective function, as demonstrated by Ref. 28. One apparent challenge is how to get the generated structures into their mechanical and dynamical equilibria, entailing further exploration such as relaxation using the machine learning interatomic potential.[39]

## Methods

### Data for generative adversarial network

The Bi-Se database is based on the substitution of all binaries of MP, in the spirit of high-throughput calculation, previous research has selected 10981 structures of them by eliminating the large unit cell, *i.e.* only select structures with the maximum number of atoms in unit cell less than 20 and the maximum length of unit cell smaller than 10 Angstrom, and we do the substitution based on their result [27,29]. Following this criteria, several stable phases like BiSe, $Bi_4Se_3$, $Bi_8Se_7$ close to the convex hull are screened out. We use the high-throughput environment to optimize the structure and determine the thermodynamical stability of Bi-Se database and the generated structures [40], where thermodynamical stability is evaluated by calculation of the formation energy. All data in Bi-Se database are considered to calculate the convex hull. For VASP setting [41], Perdew−Burke−Ernzerhof (PBE) is adopted, the energy cutoff is 500 eV, and K space density is 40 on each direction for the Brillouin zone. The Bi-Se database is in Supplementary Data 1.

**Continuous representation for crystal structures**

Specifically, 3D voxel grids are used for a typical binary compound: two grids to record the atomic positions of two elements separately and the third one to store the lattice constants, *i.e.,* the lengths and angles of/between them. Hereafter such grids are referred as site voxel and lattice voxel, respectively. Using the probability density based on Gaussian functions, the lattice voxel is defined as

$$p_{x,y,z} = \mathrm{e}^{-\frac{\mathbf{r}_{x,y,z}^2}{2\sigma^2}}, \tag{3}$$

where *x, y, z* denotes the grid index, *p* is the probability density of the denoted grid, **r** is the real space distance between this grid to the center of unit cell, and *σ* denotes the standard deviation of the Gaussian function. Similarly, sites voxel is defined as

$$p_{x,y,z} = \sum_{i=1}^{n} \mathrm{e}^{-\frac{\mathbf{r}_{n,x,y,z}^2}{2\sigma^2}}, \tag{4}$$

where *n* is number of atoms and **r** is the real space distance between this grid point and the atom. In this way, the inverse transformation is trivial for the lattice voxel while that for the sites voxel relies on the image filter technique [27].

Autoencoder is a type of artificial neural network which is able to encode inputs into low dimension vectors, as well as decode the vectors back [42]. In this work, the autoencoder is applied to encode 3D voxel data into one-dimensional (1D) vectors in the latent space, which is realized based on 3D convolutional neural network (CNN) autoencoder. The loss function of autoencoder consists of two parts: the first is the loss of information, *i.e.,* the difference between inputs and outputs, and the second is the regulation term to prevent over fitting, which yields

$$\mathrm{loss} = \left\| \mathbf{x} - \mathbf{y} \right\|^2 + \frac{\lambda}{2} \cdot \sum \boldsymbol{\omega}^2, \tag{5}$$

where **x** and **y** are the input and output of the autoencoder, *λ* is regulation coefficient, and **ω** is weight in the autoencoder [17]. The construction of autoencoder for both site and lattice voxel are the same, where both are trained using a 90%/10% training/test set ratio. Both site and lattice voxel autoencoders have high reconstruction ratio. All of them are considered to be identical by the crystal structure comparison subroutine available in pymatgen [43], with fractional length tolerance being 0.2, sites tolerance 0.3, and angle tolerance 5 degrees.

For data transformation between real space and voxel space, the voxel used for atomic positions is 64×64×64, while the voxel for lattice parameters is 32×32×32 and the $\sigma$ used in Gaussian function is always 0.15, function "gaussian_filter" from scipy package is used. [27] The autoencoders constructed by tensorflow in python. The sites autoencoder consists of 10 3D convolutional layers, *i.e.* 5 for encoder and 5 for decoder, strides except the first and last on are 2×2×2 and activation function is leakyRELU, Adam optimizer is used, learning rate is 0.003, $\lambda$ is 0.000001, $\beta_1$ is 0.9 and $\beta_2$ is 0.99 [27]. Similar, the lattice autoencoder consists of only 8 3D convolutional layers (4 for encoder and 4 for decoder), other parameters are same to the previous one [27]. Detailed design of these two models can be found in Supplementary Table 1 and 2.

**Training of DCGAN**

The left half of Fig. 2 (a) illustrates our implementation of DCGAN to generate crystal structures. Firstly, both the generated 2D crystal graphs by the generator and original 2D crystal graphs are fed into the discriminator, which is trained to distinguish such graphs. Afterwards the generator is further trained by the feedback from the discriminator through back propagation to generate structures more similar to the original structures. By repeating the processes, the generator will evolve to be able to eventually generate structures that are indistinguishable (by the discriminator) to the original structures. Details to construct DCGAN can be found in Methods and Supplementary Information.

The DCGAN model consists of generator and discriminator, they both use Adam optimizer with 0.002 learning rate, $\beta_1$ is 0.5 and $\beta_2$ is 0.999. 2D convolutional layer, dropout and batch normalization are used in both model, while RELU, leakyRELU, tanh and sigmoid are used as activation functions. 1,000,000 steps are trained for this model for 16 hours on Quadro P2000 GPU. Design of generator and discriminator are listed in Supplementary Table 3 and 4 respectively. Latent space of the GAN model has 200 dimensions.

**Training of constraint model**

The constraint model consists of 4 convolutional 2D layer and 6 fully connected layer, conducted by keras [44]. It also uses Adam optimizer, learning rate 0.002, $\beta_1$ is 0.5 and $\beta_2$ is 0.999, optimization loss is mean square error, activation function is leakyRELU. To prevent overfitting, dropout and batch normalization are used after each convolutional layer [45]. Detailed design of this model is described in Supplementary Table 5.

**Training of CCDCGAN**

The generator, discriminator and constraint have exactly the same design as the previous model, all parameters are same as parameters in DCGAN. The only difference is the optimization objective becomes Eq. (2), where the weight of the formation energy loss is 0.1. The training time is about 18 hours under the same condition as DCGAN. Typical generated structures are listed in Supplementary Table 6, more data is available in Supplementary Data 2.

## Data Availability statement

All data needed to produce the work are available from the corresponding author.

## Code Availability statement

CCDCGAN code is available upon reasonable request from the corresponding author.

## Acknowledgements

The authors gratefully acknowledge computational time on the Lichtenberg High Performance Supercomputer. Teng Long thanks the financial support from the China Scholarship Council (CSC). Part of this work was supported by the European Research Council (ERC) under the European Union's Horizon 2020 research and innovation programme (Grant No. 743116-project Cool Innov). This work was also supported by the Deutsche Forschungsgemeinschaft (DFG, German Research Foundation) – Project-ID 405553726 – TRR 270. We also acknowledge support by the Deutsche Forschungsgemeinschaft (DFG – German Research Foundation) and the Open Access Publishing Fund of Technical University of Darmstadt.

## Author Contributions

This work originated from the discussion of H.Z., T.L. and Y.Z.. H.Z. and O.G. supervised the research. T.L. worked on the machine learning model. T.L., N.F., I.O. and C.S. worked on the DFT calculations. T.L., I.S. and Y.Z. worked on data analysis. All authors contributed in the writing.

## Competing Interests

The authors declare that they have no competing interests.